\documentclass[a4paper,11pt]{article}
\usepackage{jcappub} 
\usepackage{comment}
\usepackage{enumitem}
\usepackage{graphicx,subcaption}
\usepackage{listings}
\usepackage{amsfonts}
\usepackage{xspace}

\title{\boldmath {Exploring the detection of AQNs in large liquid detectors}}
\author{I. Lazanu}
\author[1]{and M. Parvu\note{Corresponding author.}}
\affiliation{University of Bucharest, Faculty of Physics, POBox MG-11 Bucharest-Magurele, Romania}

\emailAdd{ionel.lazanu@g.unibuc.ro, mihaela.parvu@unibuc.ro}

\abstract{Recent work from the last years has raised the possibility that a portion of Dark Matter could consist of exotic particles, such as axion (anti)quark nuggets (AQN, A$\bar{\mathrm{Q}}$N). After a brief review outlining the main features of axion antiquark nuggets, we explore potential experimental signatures that can be leveraged to search for these stable supermassive particles in future surface and underground experiments using large liquid detectors. These expected signals are discussed in relation with the specific characteristics of each detection system.}

\begin{document}

\maketitle

\flushbottom

\section{Introduction}
\label{sec:intro}

Presently the existence of dark matter is strongly supported, but its nature remains elusive. At the end of the 70's, Witten \cite{Witten:1984rs} predicted the existence of a new, and more stable state of matter in comparison to the normal nuclear matter, consisting of roughly equal numbers of up, down, and strange quarks. This idea presents many problems because the QCD phase transition is not of the first order, but rather a crossover and fast evaporation is necessary. These difficulties were avoided by Ariel Zhitnitsky considering the existence of axions in the structure of nuggets. Nuggets and anti-nuggets are composite objects with (anti)baryons in the color superconducting phase squeezed by the axion domain wall as the shell. Ariel Zhitnitsky introduced \cite{Zhitnitsky:2002qa} Axion (anti)Quark Nuggets (AQNs/A$\mathrm{\bar{Q}}$Ns) as a new type of candidate for dark matter, as an alternative to WIMPs, strangelets or nuclearites.

In the Big Bang model, it is generally assumed that the Universe began as a symmetric state with zero global baryonic charge. Later, following the scenarios of Sakharov \cite{Sakharov:1967dj} which is a set of three necessary conditions (1. violation of baryon number, 2. violation of both charge conjugation symmetry (C) and charge conjugation-parity symmetry (CP), and 3. the process must not be in thermal equilibrium), this evolved into a state with a net positive baryon number. Alternatively, in Zhitnitsky’s model of the AQN, the unobserved antibaryons come to comprise the dark matter. CP-violating processes associated with the axion $\theta$ term in QCD result in the preferential formation of antinuggets, thus: 
\begin{equation}
    B_{\text{visible}}:B_{\text{nuggets}}:B_{\text{antinuggets}} \approx 1:2:3
\end{equation}
and the net baryonic number is conserved:
\begin{equation}
    B_{\text{tot}}=0=B_{\text{nuggets}}-B_{\text{antinuggets}}+B_{\text{visible}}
\end{equation}
if $B_{\text{DM}} = B_{\text{nugget}} + B_{\text{antinugget}} = 5 B_{\text{visible}}$. The ratio $B_{\text{nugget}}/B_{\text{antinugget}} = 2/3$ is determined by CP violating parameter $\theta \sim 1$ \cite{Lawson:2013bya}. It is worth mentioning that A$\bar{\mathrm{Q}}$N can be seen as a "macroscopic form of dark matter" that does not contradict the basic available constraints from BBN and CMB \cite{SinghSidhu:2020cxw}.

\section{Overview of the AQN model}

\subsection{The structure of AQN and A$\bar{\text{Q}}$N}

Regarding the structure, inside the A(anti)QN exists more regions with distinct length scales. The central core represents the size of the nugget filled by antiquarks in the Color State (CS) phase with interactions dominated by QCD theory. Following the predictions of this theory, it is expected that the characteristic length scale $R_{\text{QCD}}$ is approximately $10^{-13}$ cm. This length is further extended in accord with the baryonic number as $R = R_{\text{QCD}} B^{1/3}$. This region is surrounded by two other regions: one represented by the presence of positrons captured by the antiquarks core (electrosphere) extended up to distances of order of $10^{-8}$ cm, comparable with the atom dimensions, and the second, corresponding to the axionic domain wall (DW), much larger, with a radius of the order $R_{\text{DW}} \approx 10^{-5}$ cm.
These AQNs carry a substantial (anti)baryon charge \( \lvert B \rvert \approx 10^3 \div 10^{33} \) \cite{Gorham:2015rfa,Forbes:2008uf}. The region with baryon numbers between \(10^3\) and \(10^{23-24}\) is typically constrained by the expected dark matter (DM) flux, and thus, the AQN flux. These nuggets exhibit an approximate nuclear density, with a typical nuclear baryon number density on the order of \(10^{40}\) cm\(^{-3}\). With these parameters the effective interaction is very small, $\sigma_{\text{AQN}}/M_{\text{AQN}} \approx (\pi R^2)/(m_p \lvert B \rvert ) \sim 10^{-10}$ cm $^2$/g. In the cited article it is supposed that the (anti)baryonic matter forms a color superconducting fluid with a gap of the order of 100 MeV.
The electric charge of the A(anti)QN core depends on the phase of the antiquarks in the color-superconducting phase, although there may be many different phases in the color-superconducting quark matter. In this structure, the electric charge in the color–flavor locking (CFL) phase is very small and can be estimated as $Z \approx -0.3 \times B^{2/3} \approx -30 \div -3 \times 10^{15}$ for $B$ between $10^3$ and $10^{24}$. This interval of values is very different in the case when phase 2CS is considered. In this case $Z \approx -5 \times 10^{-3} \times B \approx -5 \div -5 \times 10^{21}$ for an identical interval of the baryon number.

In QCD, if one of the quarks has a non-zero mass, the Lagrangian contains a term that violates the CP symmetry. Experimentally, QCD does not violate CP symmetry and this aspect is solved considering a physical field, the axion, with a potential $V(\theta)$, that goes through a phase transition at a specific temperature $T_c$, that can naturally absorb the QCD CP violating term. At $T < T_c$ the symmetry is broken. At $T \sim \Lambda_{\text{QCD}}$ we have $\theta_{\text{eff}} \to 0$ (solves CP problem) and the axion field gets a mass. The axion field interacts with photons.
After the inflation, $\theta$ oscillates before the QCD phase transition at the temperature $T_c \sim 170$ MeV, the vacuum energy remains the same everywhere, but the phase can vary and form topological defects because there must be continuity in the $[0,2\pi]$. Axion domain wall topological defects start forming; quarks and anti-quarks are trapped inside the domain walls that will form nuggets and anti-nuggets. The process of domain wall formation stops at some temperature, suggested to be $\sim$ 40 MeV. Domain walls shrink until they are stopped by nuclear Fermi pressure of the Color Superconducting phase. In this model, the free parameter is the axion mass. 

The axions play a key role in the structure and stability of AQN because they generate additional pressure to stabilize the nugget. In accord with Fischer, Liang et. al. \cite{Fischer:2018niu} the total energy of an AQN is minimized when the axion contributes one-third of its total mass.
The dynamics of nuggets are governed by Equation 12 from Reference \cite{Ge:2019voa} which depends on the effective domain wall tension, the pressure difference inside and outside the nugget, and the viscosity term which describes the friction for the domain wall bubble oscillation in plasma.

\subsection{Relic abundance of A$\bar{\mathrm{Q}}$N }

Cosmological production of A$\bar{\mathrm{Q}}$N in the early Universe allows the estimation of their relic abundance. The two necessary conditions: a nonzero dark baryon number asymmetry and a first-order phase transition can easily be satisfied for many asymmetric dark matter models and QCD-like gauge theories. Such models were developed by \cite{Ge:2019voa} and \cite{Bai:2018dxf}. At a temperature around 40 MeV the A$\bar{\mathrm{Q}}$N is completed and the next step is to establish the electrical neutrality. In the case of nuggets composed of matter, these will collect electrons on the electrosphere, but this component does not essentially modify their mass. For nuggets composed of antimatter, the processes are more complicated and details were discussed by Ge and others in the cited paper.

The number of AQNs hitting the earth's surface is tiny and can be estimated in accord with \cite{Liang:2019lya}: 

\begin{equation}
 \varnothing_{\text{axions}}^{\text{Earth}} = \frac{\langle \dot{N} \rangle}{4\pi R_{\text{Earth}} ^2} = 0.4 \times \frac{10^{24}}{\langle B \rangle} \frac{\rho_{\text{DM}}}{0.3 \, \text{GeV/cm}^3} \frac{\langle v_{\text{AQN}} \rangle}{220 \, \text{km/s}} \left[ \frac{1}{\text{km}^2 \, \text{yr}} \right]
\end{equation}
which suggests that lighter AQN have a higher total flux. Considering the local dark matter with an average velocity $<v_{\text{AQN}}>$, and the density values ($\rho_{\text{DM}}$) in the range $\sim 0.3 - 0.4$ GeV/cm$^3$ \cite{deSalas:2019pee}, the AQN flux on Earth ($\varnothing_{\text{axions}}^{\text{Earth}}$) is estimated to be between $4 \times 10^3 \div 0.04$  AQN/(km$^2$ yr) for $<B>=10^{20} \div 10^{25}$.
If the AQNs are stopped inside the Earth and finally annihilate in the Earth's core, following the formalism discussed previously, it is possible to estimate the corresponding axion flux as a function of axion mass:
\begin{equation}
m_{\text{axion}} \varnothing_{\text{axions}}^{\text{Earth}} = \frac{2}{3} B[\text{GeV}] \times \varnothing_{\text{axions}}^{\text{Earth}} \left[ \frac{1}{\text{km}^2 \, \text{yr}} \right]
\end{equation}

In principle, it can be supposed that the full AQN annihilates in this process. For axions with masses of the order of $10^{-6} \mathrm{eV} \le m_a \le 10^{-3} \mathrm{eV}$ \cite{Zhitnitsky:2002qa} their fluxes are in the range of $10^{35} \div 10^{38}$ [1/(km$^2$ yr)].

\subsection{Interaction rates of A$\bar{\mathrm{Q}}$N in the atmosphere and inside Earth}

In simplified hypotheses, the energy-loss mechanism occurs through atomic collisions. When the dimensions of a nugget are larger than the atomic dimensions, its cross section can be considered as geometric, contrary, the effective area is determined by the electronic components. The energy losses of the AQN/A$\bar{\mathrm{Q}}$N along the trajectories, can be calculated following the formalism of De Rujula and Glashow \cite{DeRujula:1984axn} for nuclearites:
\begin{equation}
    -\frac{dE}{dx} = \begin{cases}
\sigma_{\text{AQN}}\rho(x) v_{\text{AQN}}^2 & \text{if } v_{\text{AQN}}(L) \geq \sqrt{\frac{\varepsilon}{\rho}} \\
\varepsilon \sigma_{\text{AQN}} & \text{if } v_{\text{AQN}}(L) < \sqrt{\frac{\varepsilon}{\rho}}
\end{cases}
\end{equation}
where $\sigma_{\mathrm{AQN}}$ is the cross-section associated with the area of the nugget, and $\rho(x)$ is the density of the medium. 
For a well-defined distance $L$, 
\begin{equation}
    v_{\text{AQN}}(L) = v_{\text{AQN}}(0) \exp \left( -\frac{\sigma_{\text{AQN}}}{M_{\text{AQN}}} \int_0^L \rho \, dx \right)
\end{equation}
where the cross section can be modeled as $\sigma_{\text{AQN}} = \pi R_{\text{AQN}}^2, \quad \text{and} \quad R_{\text{AQN}} = 10^{-7} \left( \frac{B}{10^{24}} \right)^{1/3} \, \text{m}$.

The equation of the energy loss breaks down at low velocity when the force generated by the particle becomes equal or lower than the force by which the material is confined. For velocities below the value $v_{\text{AQN}}(L) < \sqrt{\frac{\varepsilon}{\rho}}$ the energy loss decreases with a constant rate and is brought to zero.
The problem of A(anti)QN’s impacting the Earth’s atmosphere and lithosphere was investigated by Zhitnitsky and Gorham \cite{Zhitnitsky:2018mav} and \cite{Gorham:2012hy}. 

The parameters of interest for the atmosphere vary strongly depending on a multitude of factors. The average variation of air pressure (P), temperature (T), and density ($\rho$) with altitude is called the standard atmosphere and the mean values of the parameters are available \cite{engineeringtoolbox}. The atmosphere composition can be approximated considering only molecular nitrogen which represents nearly 78\% of the atmosphere concentration and oxygen 21\%. Molecular nitrogen has a triple bond between the two atoms, one sigma bond, and two pi bonds. This bond is very strong and requires 941 kJ/mol of energy to break and is only 495 kJ/mol for molecular oxygen. Considering an average density for the atmosphere around 0.657 kg/m$^3$ (as a value between sea level and $2 \times 10^4$ m), the energy loss is proportional with $\nu^2$ up to approximately 6.4 m/s. 
In the lithosphere, the integrity of rocks persists up to the energy density of $\varepsilon$=0.1 eV per molecular bond (or equivalent $\varepsilon=6.2 ×10^{20}$  eV/cm$^3$). For the processes considered inside Earth, most of the major rocks and minerals have very similar densities, around 2.6 to 3.0 g/cm$^3$ (for depths at which detection systems are usually placed). In this case, for the velocities of nuggets below the value 188 m/s the energy loss is independent of their motion. At rest, nuggets will accumulate in the Earth’s crust and annihilate. If the annihilation occurs in the vicinity of a detection system, then the effects may be detectable. A detailed discussion of these aspects will be given further from a perspective other than Gorham's \cite{Gorham:2012hy}.
. (if available), Title of the document, Year of publication (if available)
\section{Processes generated by the structure of A$\bar{\mathrm{Q}}$N and the signals of interest for direct detection}

In this section, we investigate the possibility of dark matter nuggets producing measurable signals inside the detector that are distinguishable from bias and other sources.

The spectral emissivity of photon energies below the electron mass of a nugget at effective temperature $T$, in the frame of the Thomas-Fermi model, was previously calculated \cite{Forbes:2008uf,Forbes:2009wg,Gorham:2012hy} and particular results are given for three values of the effective surface temperature (10 keV, 30 keV, 80 keV) in Figure 1 of Gorham's paper \cite{Gorham:2012hy}. 

In the present discussion, we will consider only the case of A(anti)QN composed by a core with antiquarks structure in a color superconductivity phase (2CS or CLF state), an axion domain wall, and an external electrosphere with positrons bounded to establish a neutral system. These A(anti)QN will strongly interact with other nuggets or ordinary nuclear matter. The constituent subsystems interact. A simplified hypothesis assumes a partially decoupled approach between the different processes of interaction with the external medium.

\textbf{a. Ionization processes}. For an isolated nugget, the system is neutral electric. A large number of weakly bound positrons from the electrosphere get excited and can easily leave the system. Thus, A(anti)QN acquires a negative electric charge $\approx |e|Q$, with 
\begin{equation}
    Q \approx 1.5 \times 10^6 \left( \frac{T}{\text{eV}} \right)^{\frac{5}{4}} \left( \frac{R}{2.25 \times 10^{-5} \, \text{cm}} \right)^2
\end{equation}
and can ionize the medium as a large object with mass \( M \approx m_p B \).

\textbf{b. Interaction atom-antiatom, with one in a neutral state and the other ionized.} In general, the atom-antiatom interactions are important in the treatment of annihilation processes. In this case, the problem consists in the fact that the anti-nugget structure is not in reality an antiatom, but in an approximate approach it can be considered as such. In any matter-antimatter mixture in which the atoms and antiatoms are not fully ionized, the rearrangement collisions can lead to bound states of the nucleus and antinucleus from which the annihilation can proceed. Two distinct classes of processes exist. The first corresponds to the annihilations between electrons and positrons; the second is associated with the processes between nuclei and the antinuclear core. Since the cross sections for such collisions at low energies are considerably higher than the direct particle-antiparticle annihilation cross sections, atom-antiatom interactions may play a dominant role in such a mixture. At energies lower than 100 eV, the rearrangement collisions are important, the annihilation follows the formation of the nucleus-antinucleus bound state. For the energies above this value, the direct particle-antiparticle annihilation cross-sections apply. 

The rearrangement cross section at low energies for which the interatomic potential energy is negative and proportional to the distance at a negative power, applicable for the cases in which one member is neutral and the second ionized was investigated by Morgan et. al. \cite{Morgan:1973zz}. For these cases, the cross-section can be approximated as
\begin{equation}
    \sigma = \left( \frac{2\alpha}{E} \right)^{1/2} \pi a_0^2,
\end{equation}
where $\alpha$ is the polarizability of the neutral member (medium in this case), $a_0$ is the Bohr radius ($0.529 \times 10^{-10}$ m), with $\alpha$ and $E$ in atomic units (a.u.). As an example, for Ar the polarizability is 11.083 a.u., and for Xenon 27.32 a.u. respectively \cite{doi:10.1080/00268976.2018.1535143}. With these values, considering an energy around 1 eV, the corresponding cross sections are $\sigma = 24.57 \pi a_0^2$ and $\sigma = 38.5 \pi a_0^2$ respectively.

\textbf{c. Annihilation of positrons.} In accord with Flambaum et. al. \cite{Flambaum:2021xub} the effective charge of the electron in the positron cloud varies from 1/3 in the ultrarelativistic case to 2/3 in the non-relativistic case.
Taking into account the screening of the electron charge in the positron cloud, the probability of direct annihilation is $4 \times 10^{-9}$ and $2 \times 10^{-8}$ for positronium formation probability in the 3-body collision, considering the following parameters: interaction energy 0.25 eV, and the positron density at this point: $n_{e^+} \approx 1.5 \times 10^{-4}a_0^{-3}$. 

\textbf{d. Domain wall, its properties and interactions.} In models like the one proposed by Huang and Sikivie \cite{Huang:1985tt}, the axionic domain walls appear with finite size on the order of the horizon correlated with a temperature $T_{\text{QCD}} \approx 100$ MeV and the probability of a larger size being exponentially suppressed. Sikivie \cite{Sikivie:1984yz} showed that the domain wall (DW) does not carry any electromagnetic field of its own - see Equations 4 and 5 from the cited paper, but the coupling is permitted. The strength of the coupling hold in GUT is:
\begin{equation}
    -\frac{1}{137} \frac{1}{8\pi} \frac{Na}{\langle \Phi \rangle} F_{\mu\nu} \overline{F}^{\mu\nu},
\end{equation}
where the unrenormalized value of the electroweak angle is $\sin^2 \theta_W^0 = \frac{3}{8}$ and $a$ is the axion field before mixing with the pion. For a static wall in the plane xy and a static magnetic or electric field, supposed to be oriented in the z direction, the domain wall becomes electrically charged with surface charge density:  $\frac{2e^2}{3\pi} \overrightarrow{\rm{n}} \cdot \overrightarrow{\rm{B}}$ or $-\frac{2e^2}{3\pi} \overrightarrow{\rm{n}} \times \overrightarrow{\rm{E}}$ respectively. Here, $\overrightarrow{\rm{n}}$ is the direction of increasing $\theta$. Because the environmental medium is supposed to be ionized, the conditions to generate a charge density on the surface of the DW are satisfied. The domain walls can oscillate, and dissipate energy in the form of gravitational radiation, and this process permits to estimate the lifetime independent of its size \cite{Vilenkin:1982ks,Huang:1985tt,Kibble:1980mv}. 


Budker and co-workers \cite{Budker:2019zka} suggest that the population of galactic axions with  masses of the order of $10^{-6} \mathrm{ eV} \le m_a \le 10^{-3} \mathrm{ eV}$ consists in particles with two distinct origins: a nonrelativistic component with an average velocity around $10^{-3}$c produced by different astrophysical processes, and a second relativistic component, with typical velocities 0.6c, emitted by A$\bar{\mathrm{Q}}$Ns.

Here, the aspect of interest is whether the presence of axions can generate an observable signal as an effect of the interaction. 

\textbf{Axion-photon conversion.} In accord with Di Luzio et. al. \cite{DiLuzio:2021qct}, for the coupling axion–photon $g_{a\gamma}$ and effective axion–nucleon coupling $g_{aN}^{eff}$, we are going to assume that the axion is massless or very light. If the massless case is considered, the conversion probability of an axion to a photon $P_{a\to \gamma}$ is energy independent and $P_{a\to\gamma} = \frac{g_{a \gamma}^2 B^2 L^2}{4}$, where $B$ is the magnetic field and $L$ is the length of the conversion volume. When the mass of the axion is considered, the probability must be supplemented by multiplication with the term $\frac{2(1-\cos(qL))}{(qL)^2}$, where $q$ is the transferred momentum given by $m_a^2/(2\omega)$ and $\omega$ is the energy of the axion. This probability is negligible or zero. For relativistic axions, it is supposed that they will be characterized by large dispersions in the expected signals, and thus, a broadband detector is necessary to be used \cite{Budker:2019zka}.

\textbf{Axion-electron interactions, inverse Primakoff conversion on nuclei and the axioelectric effect} \cite{Borexino:2012guz}. An axion can scatter off an electron to produce a photon in the Compton-like process $a+e \to \gamma+e$. The energy spectrum of the photons depends on the axion mass, while the spectra of electrons can be determined from $E_e=E_a-E_{\gamma}$. The last two effects are $a+Z \to \gamma+Z$ and $a+e+Z \to e+Z$. In the considered hypothesis regarding the mass of the axion, these mechanisms are not capable of producing observable signals from axion interactions.

\textbf{e. Annihilation of antiquarks from the structure of anti-nuggets and nucleons.} Quark matter occurs in various forms, depending on the temperature $T$ and quark chemical potential. At sufficiently high density and low temperature, it is possible to imagine that quarks form a degenerate Fermi liquid. Because QCD is asymptotically free, the quarks near the Fermi surface are almost free. The quark-quark interaction is certainly attractive in some channels since we know that quarks bind together to form baryons. These conditions are sufficient to guarantee color superconductivity at sufficiently high density. At zero temperature, the thermodynamic potential which we loosely refer to as the "free energy" is $E-\mu N$, where $E$ is the total energy of the system, $\mu$ is the chemical potential, and $N$ is the number of fermions. If there were no interactions, then the energy required to add or subtract particles or holes near the Fermi surface would cost zero energy. With a weak attractive interaction in any channel, if we add a pair of particles or holes with the quantum numbers of the attractive channel, the free energy is lowered by the potential energy of their attraction. Numerous pairs will be generated in the vicinity of the Fermi surface, and as they are bosonic, they will combine to form a condensate. The ground state will consist of a superposition of states with varying pair numbers, leading to the breaking of the fermion number symmetry \cite{Alford_2006}. This argument, originally developed by Bardeen, Cooper, and Schrieffer (BCS theory) is completely general in QCD: the “color Coulomb” interaction is attractive between quarks whose color wave function is antisymmetric, meaning that superconductivity arises as a direct consequence of the primary interaction in the theory \cite{Alford_2008}. The pairs of quarks cannot be color singlets and the Cooper pair condensate in quark matter will break the local color symmetry. Jiang and Kuo \cite{Jiang:1988ovi} investigated and evaluated the binding energy per nucleon for superconducting nuclear matter and compared the results with the values corresponding to normal nuclear matter. In Table \ref{Table1} the ratio for ground-state energy per nucleon, expressed in MeV, for superconducting (s) and normal (n) nuclear matter are given for different values of Fermi’s momentum in the frame of Skyrme interaction.

\begin{table}[htbp]
\begin{center}
\begin{tabular}{ |c|c|c|c|c|c|c|c| } 
 \hline
 $k_F$ [fm$^{-1}$] & 0.6 & 0.7 & 0.8 & 0.9 & 1.0 & 1.1 & 1.2 \\ 
 \hline
 $\frac{ -\left( E_0^S/N \right) }{ -\left( E_0^n/N \right) }
$ & 1.056 & 1.028 & 1.018 & 1.010 & 1.006 & 1.002 & 1.002 \\ 
 \hline
\end{tabular}
\caption{\label{Table1} Ratio for ground-state energy per nucleon, in MeV, for superconducting and normal nuclear matter for different values of Fermi’s momentum.}
\end{center}
\end{table}

Because the charge $Q$ can be large, it is possible that a nugget will capture protons from the surrounding medium considering these as quasi-free particles. Capture radius depends on both temperatures: (i) internal ($T$) corresponding to the electric charge of the AQN because of $Q(T)$ and (ii) external gas temperature – essentially determined by the hypothesis of ionized plasma as an environmental medium where the temperature ($T_{\text{gas}}$) is generated by the typical velocities of protons in plasma. The parameterization is: 
\begin{equation}
    R_{\text{capture}}(T) \approx 0.2 \, \text{cm} \times \left( \frac{T}{\text{eV}} \right)^{5/4} \left( \frac{\text{eV}}{T_{\text{gas}}} \right) \gg R_{\text{geom}}.
\end{equation}

When the AQN enters in the region of relevant baryon density, the annihilation processes start and the internal temperature increases. In these conditions, the number of annihilation events per time per unit volume is:
\begin{equation}
    \frac{dN_{\text{ann}}}{dt dV} \approx \left[ \pi R_{\text{capture}}^2(T) n n_{\text{AQN}} v_{\text{AQN}} \right] x_e (2 \, \text{GeV}).
\end{equation}
Here $n$ represents the baryon number density of the surrounding material; $n_{\text{AQN}} \approx 0.3 \times 10^{-25} (10^{25}/B) \mathrm{cm}^{-3}$ and usually it is supposed that $v_{\text{AQN}} \approx 0.3 c$. In the case of proton captures, $n$ is the proton number density. In this equation, the square bracket is the number of protons captured per time and unit volume and the total rate of annihilation is the product between this number and the fraction of ionization. If $x_e (T_{\text{gas}})$ is the ionization fraction of the gas as a function of temperature, for highly ionized gas it is possible to consider $x_e \approx 1$; for other cases $x_e \le 1$.
The temperature will be estimated using the following equation:
\begin{equation}
    T \approx 0.4 \, \text{eV} \left( \frac{n}{\text{cm}^{-3}} \right)^{4/17} \left( \frac{v_{\text{AQN}}}{10^{-3}c} \right)^{4/17} \kappa^{4/17},
\end{equation}
where $\kappa$ is a factor that considers the theoretical uncertainties in the annihilation process \cite{Zhitnitsky:2023znn} and $\kappa \le 1$. A plausible value is $\kappa \approx 0.25$ \cite{Flambaum:2023rvn}.

Unfortunately, the details of the annihilation process of baryons with antiquark matter in the color superconducting phase are not known. In order to estimate the annihilation probability for the incident antiquark core with nucleons, the single possibility is to assume that the typical cross-section of the nucleon on the antiquark core is similar to the antiproton cross-section on nuclear matter.

There have been several studies regarding the annihilation of antinucleons on nuclei. The mechanism in which antiprotons annihilate in interaction with nuclei was clearly explained by Egidy \cite{VonEgidy:1987mz}. In accord with the review of J-M Richard \cite{Richard:2019dic} a typical scenario is: a primary annihilation produces mesons, and some of them penetrate the nucleus, giving rise to a variety of phenomena: pion production, nucleon emission, internal excitation, etc. Some detailed properties have been studied. In accord with recent results \cite{Aghai-Khozani:2018hnb}, for the annihilation cross sections, if we exclude effects that may regard specific nuclei, general arguments suggest some systematic trends at large and small energies: $\sigma_{\text{ann}}$ is almost energy independent near its black-sphere value $4 \pi R_{\text{eff}}^2$ (where $R_{\text{eff}}$ is not far from the nuclear radius). In particular, one expects $\sigma_{\text{ann}}(p) \approx \sigma_{\text{ann}}(n)$, and both proportional to $A^{2/3}$. For $p \to 0$, in the absence of resonances, $\sigma_{\text{ann}} \to 1/p (n)$ and $1/p^2 (p)$. 
For a comparison with the experimental data, we used results from one classical article on nucleon–antinucleon processes \cite{PhysRev.113.1615}, which are in accord with the usual models and have been used for different interpretations of simulated data. The main experimental data is given in Table \ref{Table2}:

\begin{table}[htbp]
\begin{center}
\begin{tabular}{ |p{6cm}||p{1.5cm}|p{1.5cm}|p{4cm}|  }
 \hline
 Particles produced in annihilation in complex nuclei can be divided into: & At rest & In flight & Dominant decay mode  \\ 
 \hline
 Charged pions & $48 \pm 6 \%$ & $45 \pm 7 \%$ & $\mu\nu$ $(\approx 100\%)$ \\ 
 \hline
 Neutral particles (other than $n$, $K^0$), in particular $\pi^0$ & $28 \pm 7 \%$ & $22 \pm 7 \%$ & $\gamma \gamma$ $(\approx 98\%)$ \\ 
 \hline
 $K$ mesons & $3 \pm 1.5 \%$ & $3 \pm 1.5 \%$ & $\mu \nu (63.6\%) + \pi^{-/+} \pi^0 + 3 \pi$ (and thus $\mu \nu + \gamma \gamma$)\\ 
 \hline
Cascade of nucleons and nuclear excitation & $21 \pm 2 \%$ & $30 \pm 2 \%$ & $\gamma$\\ 
 \hline
\end{tabular}
\caption{\label{Table2} The classes of particles produced in annihilation in complex nuclei, at rest and in flight}
\end{center}
\end{table}

The primary antiproton annihilation gives rise to the number of pions (as average at rest and in flight): $<N_\pi>=5.36\pm0.3$, with average total energy (at rest and in flight): $<E_\pi>=350\pm18 \mathrm{MeV}/\pi$. Out of these pions 1.3 and 1.9 interact with the nucleus at rest and in flight respectively, giving rise to nuclear excitation and nucleon emission. 0.4 of interacting pions are inelastically scattered and the effect is a degradation of the primary pion energy to $<E_\pi>=339\pm18$ MeV. An average number of $1.6\pm0.1$ of the pions produced in the annihilation interacts with the nucleus in which the annihilation occurs, with the effect of nuclear excitation and nucleon emission. The average number of protons emitted per annihilation is $<N_B>=4.1\pm0.3$ and the corresponding total average energy release in protons and neutrons is $<\Sigma E_B>=490\pm40$ MeV. Thus, in the laboratory frame, from $\pi^0 \to \gamma + \gamma$ the energies of photons are in the energy range $0 \leq E_{\gamma} \leq E_{\pi^0}$. The distribution is flat. For the decay of charged pions, $\pi \to \mu+\nu$, the energy of muon is $0.58 E_{\pi} \leq E_{\mu} \leq E_{\pi}$. Since the annihilation processes occur within nuclei, nuclear effects may alter these results. However, the similarity between values at rest and in flight suggests minimal differences. It is expected that the annihilation process occurs throughout the entire structure of the core of nuggets formed by antiquarks.
If the annihilation process is produced in the environmental region, the only particles of interest for the identification of nuggets are the photons produced directly from the decay of neutral pions and as secondary particles from other decays. These are the only primary signals directly generated by the nuggets that can penetrate the active medium of the detection systems. The distribution of the opening angle ($\alpha$) between the two photons from $\pi^0$ in the lab frame is:
\begin{equation}
\label{photon_angle}
\frac{dN}{d\alpha} = \frac{1}{4 \beta \gamma} \cdot \frac{1}{\sqrt{\gamma^2 \sin^2\left(\frac{\alpha}{2}\right)} - 1} \cdot \frac{\cot\left(\frac{\alpha}{2}\right)}{\sin\left(\frac{\alpha}{2}\right)},
\end{equation}
peaked at the value $\alpha_{\text{min}}$ where $\sin\left(\frac{\alpha_{\text{min}}}{2}\right) = \frac{1}{\gamma}$ and vanishes at $\alpha_{\text{max}}=\pi$.

\section{Numerical results}


 
In the following, we consider different types of signals that can be produced inside the active medium of the detector. 

(i) For a simplified analysis, we can consider that the annihilation of the core region of the nuggets takes place directly in the liquid active medium of the considered detection systems. The positive and neutral pions, resulting from annihilation, are the dominant particles. As charged pions predominantly decay into muons and neutrinos, the photons produced from the decay of neutral pions serve as the primary source of measurable signals, allowing for the detection of the A$\bar{\mathrm{Q}}$Ns.
With a high probability, the signals resulting from photon interactions in the considered active media (liquid argon or liquid xenon) appear as electromagnetic showers and will be observed in coincidence. In the laboratory system, the opening angle between the primary photons is described by Equation \ref{photon_angle}.

\begin{figure}[h!]
  \centering
  \begin{subfigure}{.5\textwidth}
    \includegraphics[width=.9\linewidth]{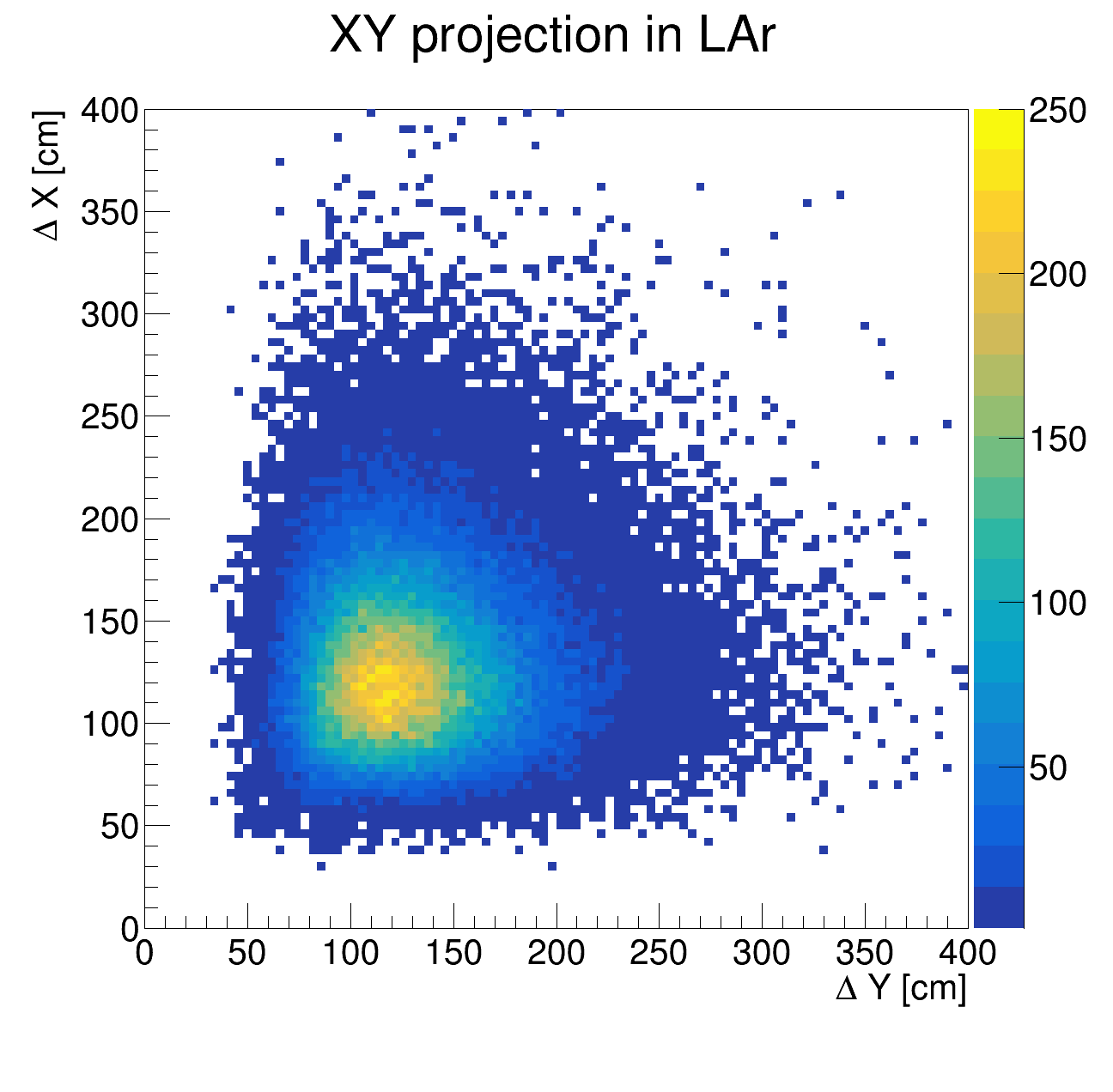}
    \label{LAr_unif_XY}
  \end{subfigure}%
  \begin{subfigure}{.5\textwidth}
    \includegraphics[width=.9\linewidth]{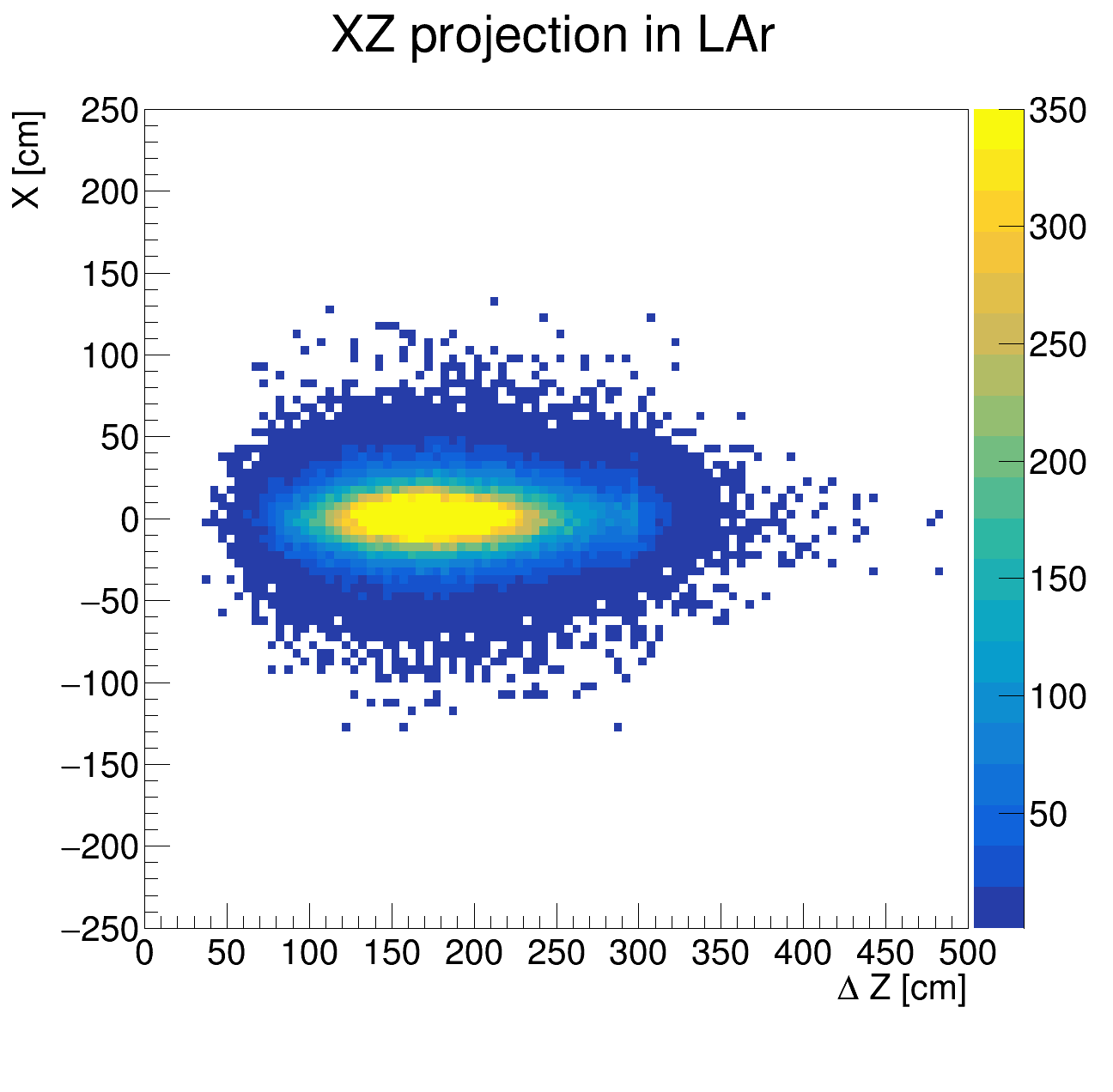}
    \label{LAr_unif_XZ}
  \end{subfigure}
  \caption{The spatial evolution of the electromagnetic cascade generated by $\pi^0$ decay in LAr considering a uniform energy distribution up to 500 MeV.}
      \label{LAr_XY_XZ}
\end{figure}

\begin{figure}[h!]
  \centering
  \begin{subfigure}{.5\textwidth}
    \includegraphics[width=.9\linewidth]{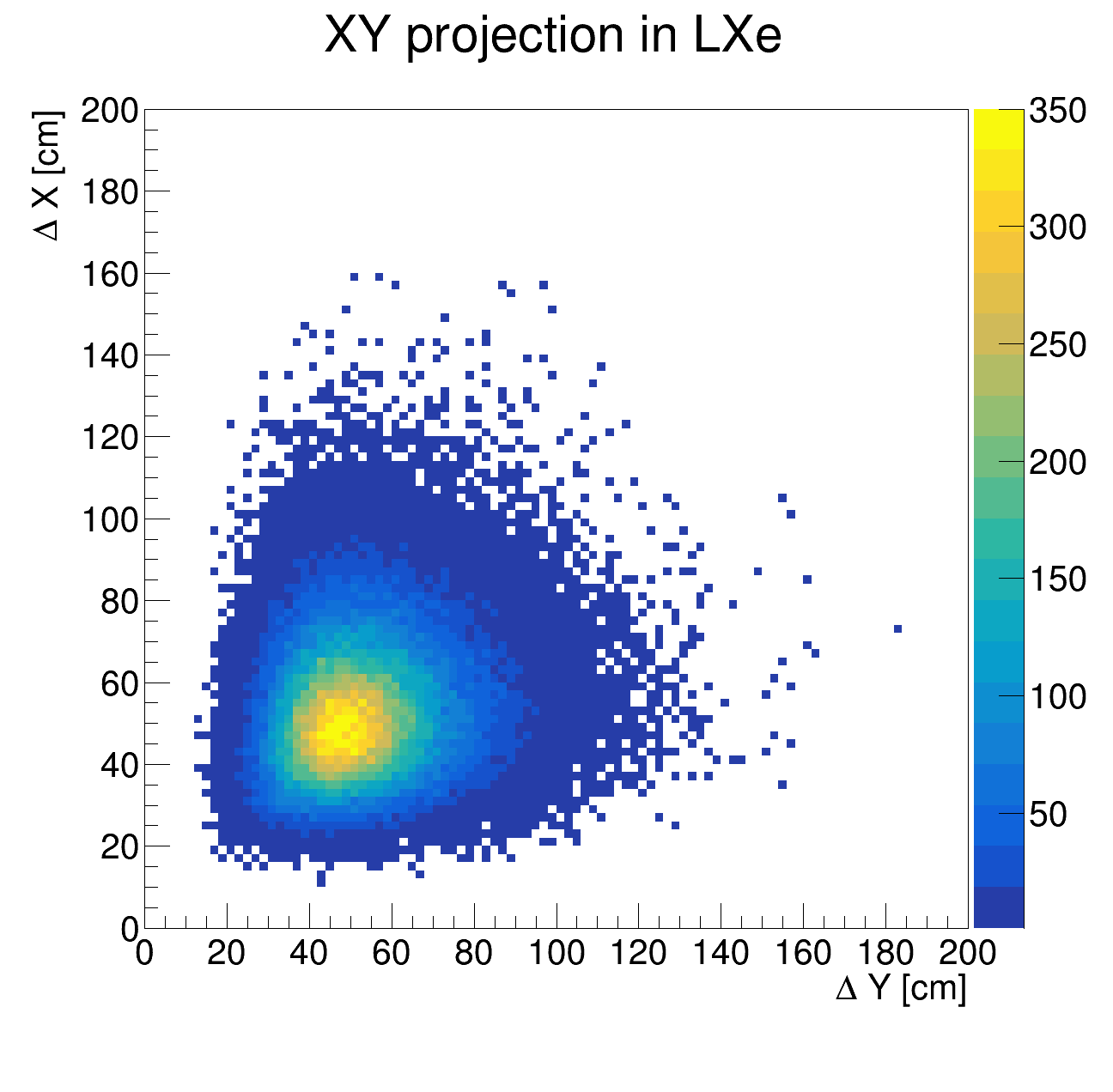}
    \label{LXe_unif_XY}
  \end{subfigure}%
  \begin{subfigure}{.5\textwidth}
    \includegraphics[width=.9\linewidth]{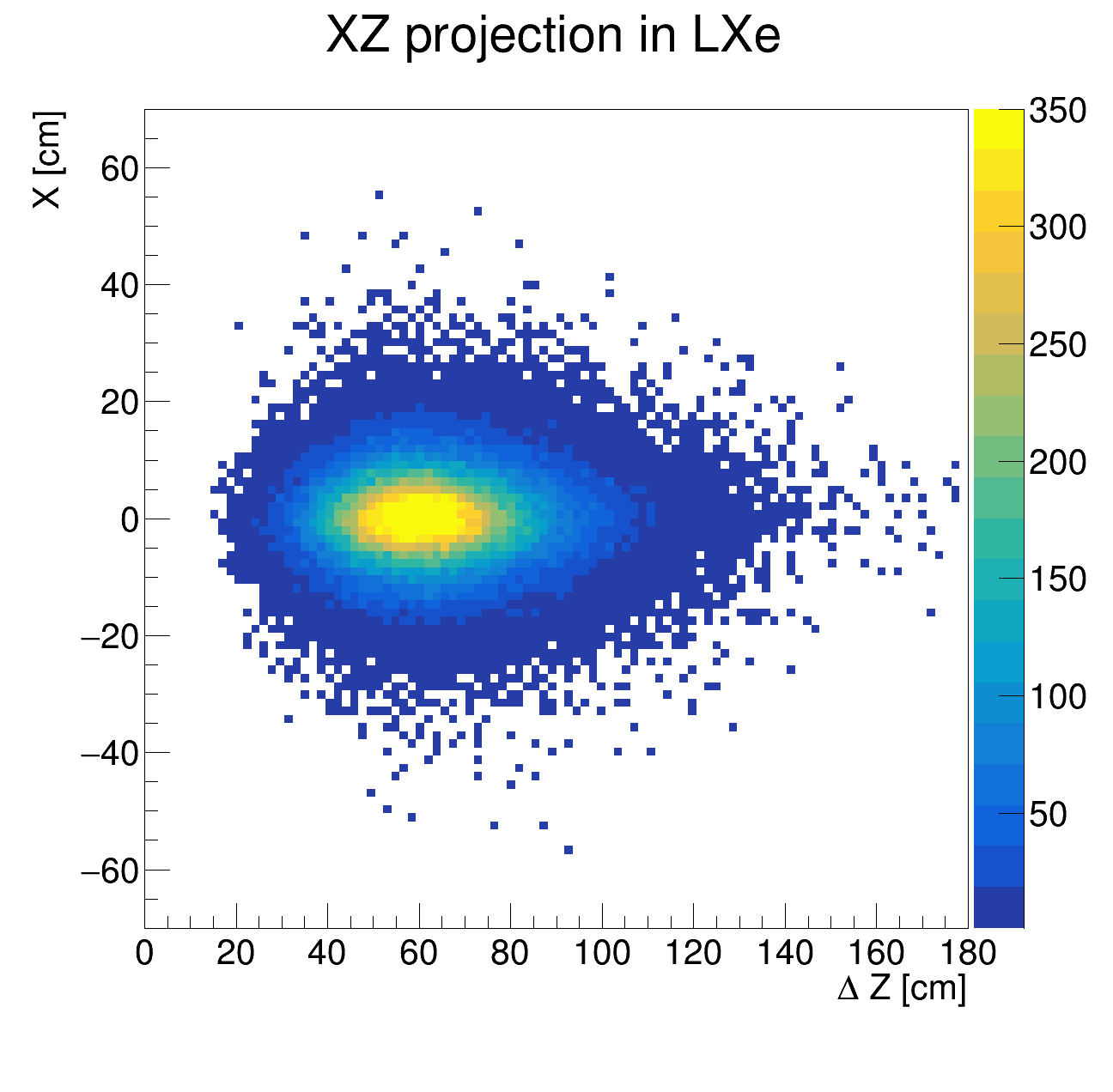}
    \label{LXe_unif_XZ}
  \end{subfigure}
  \caption{The spatial evolution of the electromagnetic cascade generated by $\pi^0$ decay in LXe considering a uniform energy distribution up to 500 MeV.}
      \label{LXe_XY_XZ}
\end{figure}

\begin{figure}[h!]
  \centering
    \includegraphics[width=.7\linewidth]{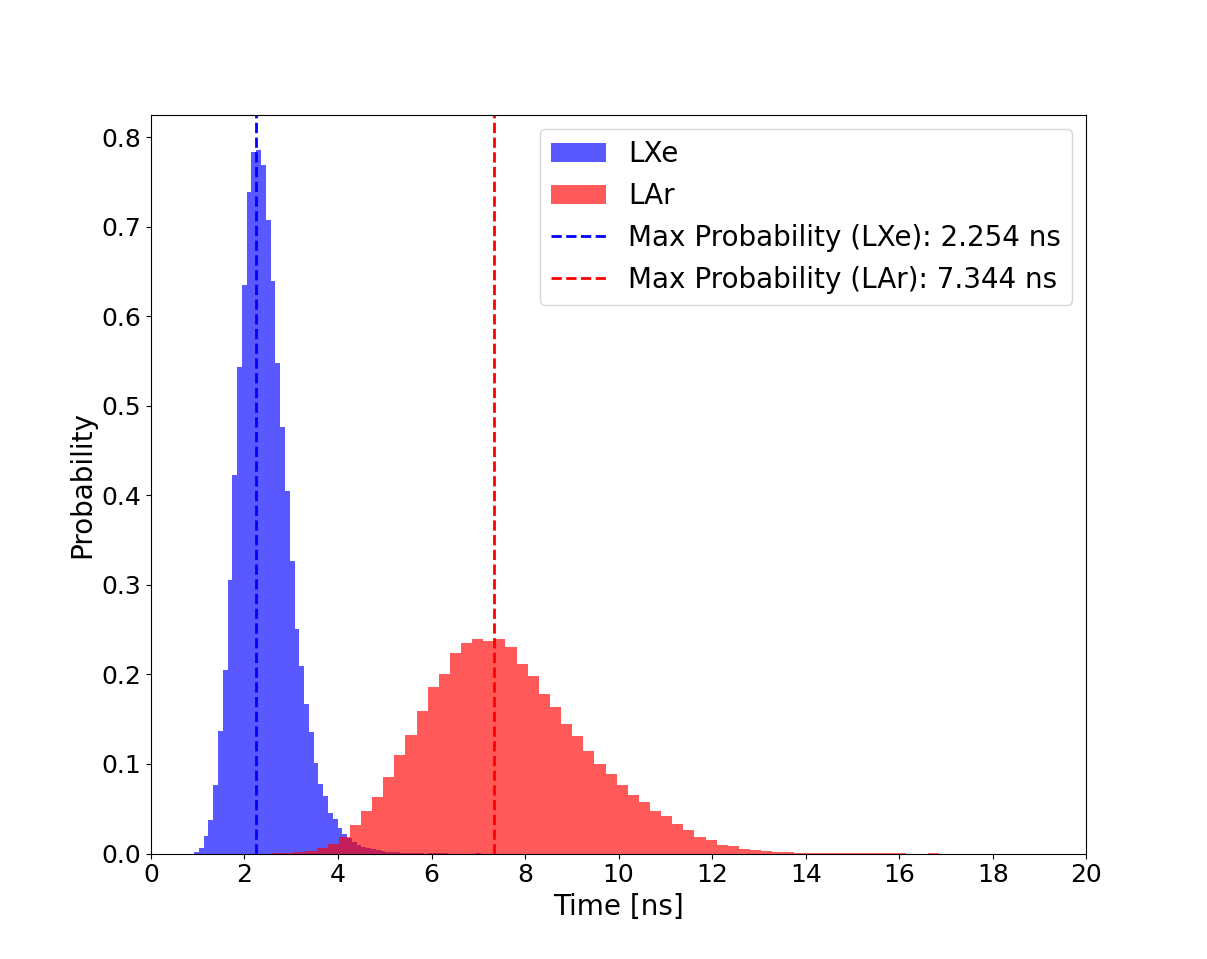}
  \caption{A comparison between the time necessary to develop the complete electromagnetic shower in LAr and LXe considering a uniform energy distribution up to 500 MeV for the decaying $\pi^0$.}
      \label{fig:comparison-time}
\end{figure}

To simulate the expected signals, we have used the FLUKA 4-3.4 code \cite{ahdida2022new,battistoni2015overview} alongside the Flair graphical interface \cite{vlachoudis2009flair}. The output has been written using the USDRAW entry of the MGDRAW subroutine and analyzed in the ROOT framework \cite{brun1997root}.
The spatial evolution of the showers induced by photons from pion decay is presented in Figures \ref{LAr_XY_XZ} and \ref{LXe_XY_XZ} for LAr and LXe. In both cases, the projections in two planes are represented. As expected, the showers develop over significantly different distances in these cases. Figure \ref{fig:comparison-time} presents a comparison between the time necessary in every case to complete the development of the shower.

A distinct problem is the capability of these detectors to discriminate between the signals of interest and the background and the identification of other sources that mimic the expected processes. Table \ref{Table3} presents the main properties of interest for the present analysis for Ar and Xe.

\begin{table}[htbp]
\begin{center}
\begin{tabular}{ |p{5.5cm}|p{1.5cm}|p{1.5cm}|  }
 \hline
Target& LAr & LXe \\ 
 \hline
Atomic number & 18 & 54 \\ 
 \hline
Atomic mass & 40 & 131.3 \\ 
 \hline
Boiling point T$_\mathrm{b}$ [K] & 87.3 & 165 \\ 
 \hline
Liquid density at T$_\mathrm{b}$ [g/cm$^3$] & 1.40 & 2.94 \\ 
 \hline 
Scintillation wavelength [nm] & 128 & 178 \\ 
 \hline
Ionization energy [eV] & 23.6 & 15.6 \\ 
 \hline  
\end{tabular}
\caption{\label{Table3}Noble Liquid Targets Parameters}
\end{center}
\end{table}

In ton-scale to multi-kton scale liquid detectors such as ProtoDUNE and DUNE (LAr) or LUX-ZEPLIN (LXe), numerous radioactive materials are present inside the detector as well as in the surrounding environment, such as the rock in the underground cavern walls. Internal radioactivity is associated not only with dust deposition and radon daughters but also with the components within the detector. Muons, neutrons, and neutrinos represent the main sources of background from cosmogenic sources for this kind of experiment. In LAr detectors \cite{Parvu:2021lrt}, the intrinsic radioactive background is dominated by the effect of the content of $^{39}$Ar in natural atmospheric argon, with a superior limit of $(8.0 \pm 0.6) \times 10^{-16}$ g/g, or $(1.01 \pm 0.08)$ Bq/kg. This undesirable isotope with the $\beta$ end-point energy of 565 keV leads to very short tracks ($\sim$ 1 mm). The background value for the energy losses is $dE/dx\approx$2.1 MeV/cm. DUNE requires an average light yield of $>20$ photoelectrons/MeV with a minimum of 0.5 photoelectron/MeV, corresponding to a photon detection efficiency (PDE) of 2.6\% and 1.3\% respectively. Caratelli and co-workers \cite{Caratelli:2022llt} extended the discussion for the next generation of experiments. For detectors placed at the surface, particularly in the case of ProtoDUNE Vertical Drift, there exists a supplementary source of background associated with its position with respect to the beam pipe \cite{Parvu:2021ezc}. In the LUX Zeplin experiment, a comprehensive analysis of the backgrounds is published in \cite{LZ:2022ysc}. The observed background rate after WIMP search criteria were applied was $(6.3\pm0.5)\times10^{-5}$ events/keVee/kg/day in the low-energy region after analysis of the contributions from alpha spectra of $^{222}$Rn, $^{218}$Po, $^{216}$Po, $^{214}$Po, and $^{212}$Po, activated $^{125}$Xe, $^{127}$Xe, $^{131\text{m}}$Xe, $^{129\text{m}}$Xe, and $^{133}$Xe. Cavern gamma-ray measurements consider contributions from $^{40}$K, $^{238}$U and $^{232}$Th. Background contributions from beta and gamma-ray sources produce flat spectra in the detector.

Showers initiated by photons up to approximately 350 MeV can have various sources: (i) due to primary cosmic photons, (ii) produced by $\pi^0$ decay (for example in a reaction such as $\bar{\nu_{\mu}}N \to \mu^+ \pi^0 N$, with threshold energy 740 MeV), or (iii) photons resulting directly from the bremsstrahlung of very energetic muons. The values of the muon antineutrino flux at the threshold, in the Kamioka location, are around $2 \times 10^2$ m$^{-2}$ s$^{-1}$ sr$^{-1}$ GeV and decrease at higher values \cite{Gaisser:2002jj}. 

Electromagnetic showers in the ICARUS TPC for $E >200$ MeV with $\gamma$ initiated from $\pi^0$ decay were reported by Behera \cite{Behera:2021iap}. These showers represent a very important background source. Following Behera's recommendations, overshielding of the detector is deemed necessary. Additionally, an anticoincidence measurement between external muons, acting as cosmogenic background after suppression, and signals originating from the shower generation inside the active medium is required. This measurement aims to put in evidence only the signals associated with axion nugget annihilation. Of course, similar problems appear in LXe detectors.
To help mitigate the cosmic ray flux, supplementary shielding is necessary for experiments at the ground level to be considered at least above the detector to stop particles before they reach the detector. For example, for the ICARUS detector, a 3 m thick concrete overburden was installed \cite{Behera:2021iap} and the results of the simulations are presented in the cited paper. Additional triggers for cosmogenic background studies are necessary to reject the contribution of muons, especially for very high-energy muons, able to produce energetic photons from bremsstrahlung inside the active volume. A simple system for cosmic-ray background rejection using an external muon counter stack, able to tag $\approx 80$\% of the cosmic rays, in a surface neutrino experiment, the MicroBooNE detector, is described in two successive papers \cite{Soleti:2016dxn,MicroBooNE:2017sup}. For example, in the process $\bar{\nu_{\mu}}N \to \mu^+ \pi^0 N$ the muon is produced inside the detector and is not counted; in this case, the neutral pion is a background for the nuggets searches. To reject these processes, a supplementary solution is necessary and a Cosmic Ray Tagging (CRT) system must be used. The idea of CRT is not new – see for example \cite{Fixsen:2000ke,Zhang:2019lzf}. The ICARUS collaboration also considered this method \cite{Heggestuen:2023mql} where the system surrounds the detector. 
An alternative solution was suggested for DUNE, presenting a module with extremely low background levels \cite{Bezerra:2023gvl}. Managing radioactive backgrounds from detector components involves enhancements in material selection, opting for materials with low radioactive concentrations. Additionally, measures such as increased neutron shielding and minimizing radon levels within the detector contribute to effective background control. The cryostat filled with atmospheric LAr contains a second cryostat enclosing underground LAr. The installation of appropriate electronics in both cryostats can allow a rejection of the background and simultaneously a coincident detection of the signal of interest.

Using timing information from the CRT hits and detected light signals, it is possible to discriminate between particle tracks either entering or exiting the detector or produced inside. A supplementary analysis is necessary to identify the particle. In particular, if the muon identification is possible, the rejection of $\bar{\nu_{\mu}}N \to \mu^+ \pi^0 N$ reaction is possible. Details of the time analysis of signals are presented by the ICARUS Collaboration in the cited paper.

(ii) During the annihilation processes, additional emissions may include weakly coupled axions, neutrinos, and X-rays. X-rays can also be generated in the de-excitation processes of the excimer states of LAr or LXe. The most important feature of the X-ray spectrum from AQN is that it is peaked in the (10–50) keV energy range, as claimed by Liang and co-workers \cite{Liang:2020mnz}. The interaction of photons with matter involves three competing processes: the photoelectric effect, Compton scattering, and pair production. In all these processes, electrons are produced, which lose their energy through ionization or excitation, leading to exciton production. Using photon cross sections database for each interaction process and attenuation coefficients for $\gamma$-rays in LAr \cite{berger1998xcom}, it is observed that photoelectric absorption is the predominant process below 80 keV, while above this energy, Compton scattering becomes the dominating interaction. Assuming for LAr a density of 1.396 g/cm$^{3}$ at 1 atm, the interaction length for $\gamma$-rays at 60 keV is $\sim$1.5 cm and $\sim$12.4 cm at 1 MeV \cite{Creus:2013sau}. For example, for the measurements of the low energy gamma rays, the next project GRAMS \cite{Aramaki:2019bpi}, using a large LArTPC propose methods to register and reconstruct these signals. For energies above the threshold for electron-positron production, will be possible to reconstruct the momentum of the pair and to determine the incident gamma ray. For lower energies, where photo-absorption and Compton scattering are dominant, the detector relies on accurately determining the position and energy of the electron.

(iii) The existence of the coupling of the axion to the electromagnetic field opens the window to the direct detection of axions. All rare gas atoms form fairly stable positive molecule-ions. By the addition of an electron in one or another Rydberg molecular orbital, any rare-gas molecule-ion should give rise to very numerous stable excimeric states \cite{10.1063/1.1672756}. In argon, the basic aspects are the following: light yield $\sim$ a few 10000’s of photons per MeV (depending on the electric field, particle type, and purity); the dominant wavelength of emission is 128 nm and other transitions are discussed in \cite{Parvu:2017xde}; the light with two characteristic time constants: fast component (6 ns) and slow component (1500 ns); the argon is highly transparent to its scintillation light. There are two low-lying excited states: a singlet state $1\Sigma u^+$ and a triplet state $3\Sigma u^+$. The singlet and triplet refer to how the spin of the electron and argon dimer couple in the Rydberg “atom”. The mechanisms of scintillation in LAr are self-trapped exciton luminescence and recombination luminescence \cite{10.1063/1.1672756}. For xenon, most of the details are discussed by Mulliken \cite{10.1063/1.1672756}.

The ratios between the intensities of different light transitions are known. The existence of the axion-photon coupling allows the conversion of axions and thus the excitation of higher Rydberg states. If the de-excitations are spectrometrically recorded with satisfactory precision, the change in the ratios between their intensities will be correlated with the processes induced by the axions.

\section{Summary}
This paper explores various types of signals expected to be produced as an effect of the interactions of the Axion (anti)Quark Nuggets in large liquid detectors. The expected signals are discussed in correlation with the potential for discrimination against various background sources within the active medium, along with methods to reject signals originating from cosmic rays. The primary signal types include electromagnetic showers resulting from the annihilation of the core of antiquark nuggets, X-ray spectra, and the potential for direct detection of axions through the de-excitations of the excimeric states of argon and xenon.

\acknowledgments

We thank Professor Konstantin Zioutas for the suggestion to address this topic. MP would also like to thank Nicusor Arsene (ISS) for the useful discussions regarding this subject. This work was performed with the financial support of the Romanian Program PNCDI III, Programme 5, Module 5.2 CERN-RO, under contract no. 04/2022.


\bibliographystyle{JHEP}
\bibliography{main.bib}

\providecommand{\href}[2]{#2}\begingroup\raggedright\begin{thebibliography}{10}

\bibitem{Witten:1984rs}
E.~Witten, \emph{{Cosmic Separation of Phases}}, \href{https://doi.org/10.1103/PhysRevD.30.272}{\emph{Phys. Rev. D} {\bfseries 30} (1984) 272}.

\bibitem{Zhitnitsky:2002qa}
A.R.~Zhitnitsky, \emph{{'Nonbaryonic' dark matter as baryonic color superconductor}}, \href{https://doi.org/10.1088/1475-7516/2003/10/010}{\emph{JCAP} {\bfseries 10} (2003) 010} [\href{https://arxiv.org/abs/hep-ph/0202161}{{\ttfamily hep-ph/0202161}}].

\bibitem{Sakharov:1967dj}
A.D.~Sakharov, \emph{{Violation of CP Invariance, C asymmetry, and baryon asymmetry of the universe}}, \href{https://doi.org/10.1070/PU1991v034n05ABEH002497}{\emph{Pisma Zh. Eksp. Teor. Fiz.} {\bfseries 5} (1967) 32}.

\bibitem{Lawson:2013bya}
K.~Lawson and A.R.~Zhitnitsky, \emph{{Quark (Anti) Nugget Dark Matter}},  in \emph{{Snowmass 2013}: {Snowmass on the Mississippi}}, 5, 2013 [\href{https://arxiv.org/abs/1305.6318}{{\ttfamily 1305.6318}}].

\bibitem{SinghSidhu:2020cxw}
J.~Singh~Sidhu, R.J.~Scherrer and G.~Starkman, \emph{{Antimatter as macroscopic dark matter}}, \href{https://doi.org/10.1016/j.physletb.2020.135574}{\emph{Phys. Lett. B} {\bfseries 807} (2020) 135574} [\href{https://arxiv.org/abs/2006.01200}{{\ttfamily 2006.01200}}].

\bibitem{Gorham:2015rfa}
P.W.~Gorham and B.J.~Rotter, \emph{{Stringent neutrino flux constraints on antiquark nugget dark matter}}, \href{https://doi.org/10.1103/PhysRevD.95.103002}{\emph{Phys. Rev. D} {\bfseries 95} (2017) 103002} [\href{https://arxiv.org/abs/1507.03545}{{\ttfamily 1507.03545}}].

\bibitem{Forbes:2008uf}
M.M.~Forbes and A.R.~Zhitnitsky, \emph{{WMAP Haze: Directly Observing Dark Matter?}}, \href{https://doi.org/10.1103/PhysRevD.78.083505}{\emph{Phys. Rev. D} {\bfseries 78} (2008) 083505} [\href{https://arxiv.org/abs/0802.3830}{{\ttfamily 0802.3830}}].

\bibitem{Fischer:2018niu}
H.~Fischer, X.~Liang, Y.~Semertzidis, A.~Zhitnitsky and K.~Zioutas, \emph{{New mechanism producing axions in the AQN model and how the CAST can discover them}}, \href{https://doi.org/10.1103/PhysRevD.98.043013}{\emph{Phys. Rev. D} {\bfseries 98} (2018) 043013} [\href{https://arxiv.org/abs/1805.05184}{{\ttfamily 1805.05184}}].

\bibitem{Ge:2019voa}
S.~Ge, K.~Lawson and A.~Zhitnitsky, \emph{{Axion quark nugget dark matter model: Size distribution and survival pattern}}, \href{https://doi.org/10.1103/PhysRevD.99.116017}{\emph{Phys. Rev. D} {\bfseries 99} (2019) 116017} [\href{https://arxiv.org/abs/1903.05090}{{\ttfamily 1903.05090}}].

\bibitem{Bai:2018dxf}
Y.~Bai, A.J.~Long and S.~Lu, \emph{{Dark Quark Nuggets}}, \href{https://doi.org/10.1103/PhysRevD.99.055047}{\emph{Phys. Rev. D} {\bfseries 99} (2019) 055047} [\href{https://arxiv.org/abs/1810.04360}{{\ttfamily 1810.04360}}].

\bibitem{Liang:2019lya}
X.~Liang, A.~Mead, M.S.R.~Siddiqui, L.~Van~Waerbeke and A.~Zhitnitsky, \emph{{Axion Quark Nugget Dark Matter: Time Modulations and Amplifications}}, \href{https://doi.org/10.1103/PhysRevD.101.043512}{\emph{Phys. Rev. D} {\bfseries 101} (2020) 043512} [\href{https://arxiv.org/abs/1908.04675}{{\ttfamily 1908.04675}}].

\bibitem{deSalas:2019pee}
P.F.~de~Salas, K.~Malhan, K.~Freese, K.~Hattori and M.~Valluri, \emph{{On the estimation of the Local Dark Matter Density using the rotation curve of the Milky Way}}, \href{https://doi.org/10.1088/1475-7516/2019/10/037}{\emph{JCAP} {\bfseries 10} (2019) 037} [\href{https://arxiv.org/abs/1906.06133}{{\ttfamily 1906.06133}}].

\bibitem{DeRujula:1984axn}
A.~De~Rujula and S.L.~Glashow, \emph{{Nuclearites: A Novel Form of Cosmic Radiation}}, \href{https://doi.org/10.1038/312734a0}{\emph{Nature} {\bfseries 312} (1984) 734}.

\bibitem{Zhitnitsky:2018mav}
A.~Zhitnitsky, \emph{{Solar Flares and the Axion Quark Nugget Dark Matter Model}}, \href{https://doi.org/10.1016/j.dark.2018.08.001}{\emph{Phys. Dark Univ.} {\bfseries 22} (2018) 1} [\href{https://arxiv.org/abs/1801.01509}{{\ttfamily 1801.01509}}].

\bibitem{Gorham:2012hy}
P.W.~Gorham, \emph{{Antiquark nuggets as dark matter: New constraints and detection prospects}}, \href{https://doi.org/10.1103/PhysRevD.86.123005}{\emph{Phys. Rev. D} {\bfseries 86} (2012) 123005} [\href{https://arxiv.org/abs/1208.3697}{{\ttfamily 1208.3697}}].

\bibitem{engineeringtoolbox}
E.~Toolbox, \emph{Standard atmosphere},  2024.

\bibitem{Forbes:2009wg}
M.M.~Forbes, K.~Lawson and A.R.~Zhitnitsky, \emph{{The Electrosphere of Macroscopic 'Quark Nuclei': A Source for Diffuse MeV Emissions from Dark Matter}}, \href{https://doi.org/10.1103/PhysRevD.82.083510}{\emph{Phys. Rev. D} {\bfseries 82} (2010) 083510} [\href{https://arxiv.org/abs/0910.4541}{{\ttfamily 0910.4541}}].

\bibitem{Morgan:1973zz}
D.L.~Morgan and V.W.~Hughes, \emph{{Atom-Antiatom Interactions}}, \href{https://doi.org/10.1103/PhysRevA.7.1811}{\emph{Phys. Rev. A} {\bfseries 7} (1973) 1811}.

\bibitem{doi:10.1080/00268976.2018.1535143}
P.~Schwerdtfeger and J.K.~Nagle, \emph{2018 table of static dipole polarizabilities of the neutral elements in the periodic table*}, \href{https://doi.org/10.1080/00268976.2018.1535143}{\emph{Molecular Physics} {\bfseries 117} (2019) 1200} [\href{https://arxiv.org/abs/https://doi.org/10.1080/00268976.2018.1535143}{{\ttfamily https://doi.org/10.1080/00268976.2018.1535143}}].

\bibitem{Flambaum:2021xub}
V.V.~Flambaum and I.B.~Samsonov, \emph{{Radiation from matter-antimatter annihilation in the quark nugget model of dark matter}}, \href{https://doi.org/10.1103/PhysRevD.104.063042}{\emph{Phys. Rev. D} {\bfseries 104} (2021) 063042} [\href{https://arxiv.org/abs/2108.00652}{{\ttfamily 2108.00652}}].

\bibitem{Huang:1985tt}
M.C.~Huang and P.~Sikivie, \emph{{The Structure of Axionic Domain Walls}}, \href{https://doi.org/10.1103/PhysRevD.32.1560}{\emph{Phys. Rev. D} {\bfseries 32} (1985) 1560}.

\bibitem{Sikivie:1984yz}
P.~Sikivie, \emph{{On the Interaction of Magnetic Monopoles With Axionic Domain Walls}}, \href{https://doi.org/10.1016/0370-2693(84)91731-3}{\emph{Phys. Lett. B} {\bfseries 137} (1984) 353}.

\bibitem{Vilenkin:1982ks}
A.~Vilenkin and A.E.~Everett, \emph{{Cosmic Strings and Domain Walls in Models with Goldstone and PseudoGoldstone Bosons}}, \href{https://doi.org/10.1103/PhysRevLett.48.1867}{\emph{Phys. Rev. Lett.} {\bfseries 48} (1982) 1867}.

\bibitem{Kibble:1980mv}
T.W.B.~Kibble, \emph{{Some Implications of a Cosmological Phase Transition}}, \href{https://doi.org/10.1016/0370-1573(80)90091-5}{\emph{Phys. Rept.} {\bfseries 67} (1980) 183}.

\bibitem{Budker:2019zka}
D.~Budker, V.V.~Flambaum, X.~Liang and A.~Zhitnitsky, \emph{{Axion Quark Nuggets and how a Global Network can discover them}}, \href{https://doi.org/10.1103/PhysRevD.101.043012}{\emph{Phys. Rev. D} {\bfseries 101} (2020) 043012} [\href{https://arxiv.org/abs/1909.09475}{{\ttfamily 1909.09475}}].

\bibitem{DiLuzio:2021qct}
L.~Di~Luzio et~al., \emph{{Probing the axion\textendash{}nucleon coupling with the next generation of~axion helioscopes}}, \href{https://doi.org/10.1140/epjc/s10052-022-10061-1}{\emph{Eur. Phys. J. C} {\bfseries 82} (2022) 120} [\href{https://arxiv.org/abs/2111.06407}{{\ttfamily 2111.06407}}].

\bibitem{Borexino:2012guz}
{\scshape Borexino} collaboration, \emph{{Search for Solar Axions Produced in $p(d,\rm{^3He})A$ Reaction with Borexino Detector}}, \href{https://doi.org/10.1103/PhysRevD.85.092003}{\emph{Phys. Rev. D} {\bfseries 85} (2012) 092003} [\href{https://arxiv.org/abs/1203.6258}{{\ttfamily 1203.6258}}].

\bibitem{Alford_2006}
M.~Alford and K.~Rajagopal, \emph{Color superconductivity in dense, but not asymptotically dense, quark matter},  in \emph{Pairing in Fermionic Systems}, p.~1–36, WORLD SCIENTIFIC (2006), \href{https://doi.org/10.1142/9789812773043_0001}{DOI}.

\bibitem{Alford_2008}
M.G.~Alford, A.~Schmitt, K.~Rajagopal and T.~Schäfer, \emph{Color superconductivity in dense quark matter}, \href{https://doi.org/10.1103/revmodphys.80.1455}{\emph{Reviews of Modern Physics} {\bfseries 80} (2008) 1455–1515}.

\bibitem{Jiang:1988ovi}
M.F.~Jiang and T.T.S.~Kuo, \emph{{Thermodynamic properties of superconducting nuclear matter}}, \href{https://doi.org/10.1016/0375-9474(88)90498-8}{\emph{Nucl. Phys. A} {\bfseries 481} (1988) 294}.

\bibitem{Zhitnitsky:2023znn}
A.~Zhitnitsky, \emph{{Structure formation paradigm and axion quark nugget dark matter model}}, \href{https://doi.org/10.1016/j.dark.2023.101217}{\emph{Phys. Dark Univ.} {\bfseries 40} (2023) 101217} [\href{https://arxiv.org/abs/2302.00010}{{\ttfamily 2302.00010}}].

\bibitem{Flambaum:2023rvn}
V.V.~Flambaum, I.B.~Samsonov and G.K.~Vong, \emph{{Possibility of antiquark nuggets detection using meteor searching radars}}, \href{https://doi.org/10.1103/PhysRevD.107.123501}{\emph{Phys. Rev. D} {\bfseries 107} (2023) 123501} [\href{https://arxiv.org/abs/2303.01697}{{\ttfamily 2303.01697}}].

\bibitem{VonEgidy:1987mz}
T.~Von~Egidy, \emph{{Interaction and Annihilation of Anti-protons and Nuclei}}, \href{https://doi.org/10.1038/328773a0}{\emph{Nature} {\bfseries 328} (1987) 773}.

\bibitem{Richard:2019dic}
J.-M.~Richard, \emph{{Antiproton physics}}, \href{https://doi.org/10.3389/fphy.2020.00006}{\emph{Front. in Phys.} {\bfseries 8} (2020) 6} [\href{https://arxiv.org/abs/1912.07385}{{\ttfamily 1912.07385}}].

\bibitem{Aghai-Khozani:2018hnb}
H.~Aghai-Khozani et~al., \emph{{Measurement of the antiproton\textendash{}nucleus annihilation cross-section at low energy}}, \href{https://doi.org/10.1016/j.nuclphysa.2018.01.001}{\emph{Nucl. Phys. A} {\bfseries 970} (2018) 366}.

\bibitem{PhysRev.113.1615}
O.~Chamberlain, G.~Goldhaber, L.~Jauneau, T.~Kalogeropoulos, E.~Segr\`e and R.~Silberberg, \emph{Antiproton-nucleon annihilation process. ii}, \href{https://doi.org/10.1103/PhysRev.113.1615}{\emph{Phys. Rev.} {\bfseries 113} (1959) 1615}.

\bibitem{ahdida2022new}
C.~Ahdida, D.~Bozzato, D.~Calzolari, F.~Cerutti, N.~Charitonidis, A.~Cimmino et~al., \emph{New capabilities of the fluka multi-purpose code}, {\emph{Frontiers in Physics} {\bfseries 9} (2022) 788253}.

\bibitem{battistoni2015overview}
G.~Battistoni, T.~Boehlen, F.~Cerutti, P.W.~Chin, L.S.~Esposito, A.~Fass{\`o} et~al., \emph{Overview of the fluka code}, {\emph{Annals of Nuclear Energy} {\bfseries 82} (2015) 10}.

\bibitem{vlachoudis2009flair}
V.~Vlachoudis et~al., \emph{Flair: a powerful but user friendly graphical interface for fluka},  in \emph{Proc. Int. Conf. on Mathematics, Computational Methods \& Reactor Physics (M\&C 2009), Saratoga Springs, New York}, vol.~176, 2009.

\bibitem{brun1997root}
R.~Brun and F.~Rademakers, \emph{Root—an object oriented data analysis framework}, {\emph{Nuclear instruments and methods in physics research section A: accelerators, spectrometers, detectors and associated equipment} {\bfseries 389} (1997) 81}.

\bibitem{Parvu:2021lrt}
M.~Parvu and I.~Lazanu, \emph{{Can strangelets be detected in a large LAr neutrino detector?}}, \href{https://doi.org/10.1088/1475-7516/2021/11/040}{\emph{JCAP} {\bfseries 11} (2021) 040} [\href{https://arxiv.org/abs/2107.05257}{{\ttfamily 2107.05257}}].

\bibitem{Caratelli:2022llt}
D.~Caratelli et~al., \emph{{Low-Energy Physics in Neutrino LArTPCs}},  \href{https://arxiv.org/abs/2203.00740}{{\ttfamily 2203.00740}}.

\bibitem{Parvu:2021ezc}
M.~Parvu and I.~Lazanu, \emph{{Radioactive background for ProtoDUNE detector}}, \href{https://doi.org/10.1088/1475-7516/2021/08/042}{\emph{JCAP} {\bfseries 08} (2021) 042} [\href{https://arxiv.org/abs/2104.10604}{{\ttfamily 2104.10604}}].

\bibitem{LZ:2022ysc}
{\scshape LZ} collaboration, \emph{{Background determination for the LUX-ZEPLIN dark matter experiment}}, \href{https://doi.org/10.1103/PhysRevD.108.012010}{\emph{Phys. Rev. D} {\bfseries 108} (2023) 012010} [\href{https://arxiv.org/abs/2211.17120}{{\ttfamily 2211.17120}}].

\bibitem{Gaisser:2002jj}
T.K.~Gaisser and M.~Honda, \emph{{Flux of atmospheric neutrinos}}, \href{https://doi.org/10.1146/annurev.nucl.52.050102.090645}{\emph{Ann. Rev. Nucl. Part. Sci.} {\bfseries 52} (2002) 153} [\href{https://arxiv.org/abs/hep-ph/0203272}{{\ttfamily hep-ph/0203272}}].

\bibitem{Behera:2021iap}
{\scshape ICARUS} collaboration, \emph{{Cosmogenic background suppression at the ICARUS using a concrete overburden}}, \href{https://doi.org/10.1088/1742-6596/2156/1/012181}{\emph{J. Phys. Conf. Ser.} {\bfseries 2156} (2021) 012181}.

\bibitem{Soleti:2016dxn}
S.R.~Soleti, \emph{{The Muon Counter System for the MicroBooNE experiment}},  in \emph{{Prospects in Neutrino Physics}}, 4, 2016 [\href{https://arxiv.org/abs/1604.07858}{{\ttfamily 1604.07858}}].

\bibitem{MicroBooNE:2017sup}
{\scshape MicroBooNE} collaboration, \emph{{Measurement of cosmic-ray reconstruction efficiencies in the MicroBooNE LArTPC using a small external cosmic-ray counter}}, \href{https://doi.org/10.1088/1748-0221/12/12/P12030}{\emph{JINST} {\bfseries 12} (2017) P12030} [\href{https://arxiv.org/abs/1707.09903}{{\ttfamily 1707.09903}}].

\bibitem{Fixsen:2000ke}
D.J.~Fixsen, J.D.~Offenberg, R.J.~Hanisch, J.C.~Mather, M.A.~Nieto-Santisteban, R.~Sengupta et~al., \emph{{Cosmic ray rejection and readout efficiency for large-area arrays}}, \href{https://doi.org/10.1086/316626}{\emph{Publ. Astron. Soc. Pac.} {\bfseries 112} (2000) 1350} [\href{https://arxiv.org/abs/astro-ph/0005486}{{\ttfamily astro-ph/0005486}}].

\bibitem{Zhang:2019lzf}
K.~Zhang and J.S.~Bloom, \emph{{deepCR: Cosmic Ray Rejection with Deep Learning}}, \href{https://doi.org/10.3847/1538-4357/ab3fa6}{\emph{Astrophys. J.} {\bfseries 889} (2020) 24} [\href{https://arxiv.org/abs/1907.09500}{{\ttfamily 1907.09500}}].

\bibitem{Heggestuen:2023mql}
{\scshape ICARUS} collaboration, \emph{{Light detection and Cosmic Rejection in the ICARUS LArTPC at Fermilab}},  \href{https://arxiv.org/abs/2312.05684}{{\ttfamily 2312.05684}}.

\bibitem{Bezerra:2023gvl}
T.~Bezerra et~al., \emph{{Large low background kTon-scale liquid argon time projection chambers}}, \href{https://doi.org/10.1088/1361-6471/acc394}{\emph{J. Phys. G} {\bfseries 50} (2023) 060502} [\href{https://arxiv.org/abs/2301.11878}{{\ttfamily 2301.11878}}].

\bibitem{Liang:2020mnz}
X.~Liang, E.~Peshkov, L.~Van~Waerbeke and A.~Zhitnitsky, \emph{{Proposed network to detect axion quark nugget dark matter}}, \href{https://doi.org/10.1103/PhysRevD.103.096001}{\emph{Phys. Rev. D} {\bfseries 103} (2021) 096001} [\href{https://arxiv.org/abs/2012.00765}{{\ttfamily 2012.00765}}].

\bibitem{berger1998xcom}
M.J.~Berger, \emph{Xcom: photon cross sections database}, {\emph{http://physics. nist. gov/PhysRefData/Xcom/Text/XCOM. html} {\bfseries 8} (1998) 3587}.

\bibitem{Creus:2013sau}
W.~Creus, \emph{{Light Yield in Liquid Argon for Dark Matter Detection}}, Ph.D. thesis, Zurich U., 2013.
\newblock 10.5167/uzh-86639.

\bibitem{Aramaki:2019bpi}
T.~Aramaki, P.~Hansson~Adrian, G.~Karagiorgi and H.~Odaka, \emph{{Dual MeV Gamma-Ray and Dark Matter Observatory - GRAMS Project}}, \href{https://doi.org/10.1016/j.astropartphys.2019.07.002}{\emph{Astropart. Phys.} {\bfseries 114} (2020) 107} [\href{https://arxiv.org/abs/1901.03430}{{\ttfamily 1901.03430}}].

\bibitem{10.1063/1.1672756}
R.S.~Mulliken, \emph{{Potential Curves of Diatomic Rare‐Gas Molecules and Their Ions, with Particular Reference to Xe2}}, \href{https://doi.org/10.1063/1.1672756}{\emph{The Journal of Chemical Physics} {\bfseries 52} (2003) 5170} [\href{https://arxiv.org/abs/https://pubs.aip.org/aip/jcp/article-pdf/52/10/5170/14734782/5170\_1\_online.pdf}{{\ttfamily https://pubs.aip.org/aip/jcp/article-pdf/52/10/5170/14734782/5170\_1\_online.pdf}}].

\bibitem{Parvu:2017xde}
M.~P\^arvu and I.~Lazanu, \emph{{Some considerations about cosmogenic production of radioactive isotopes in Ar as target for the next neutrino experiments}}, \href{https://doi.org/10.1016/j.radphyschem.2018.08.009}{\emph{Radiat. Phys. Chem.} {\bfseries 152} (2018) 129} [\href{https://arxiv.org/abs/1712.04399}{{\ttfamily 1712.04399}}].

\end{thebibliography}\endgroup


\end{document}